\documentclass{appolb}
\usepackage{epsfig}

\begin{document}
\title{Study of the near threshold $pp\to ppK^+K^-$ reaction\\
in view of the $K^+K^-$ final state interaction
\thanks{Presented at Excited QCD 2010}%
}
\author{M. Silarski, P. Moskal on behalf of the COSY-11 collaboration
\address{\begin{center}Institute of Physics, Jagiellonian University, PL-30-059 Cracow, Poland\\
                 \&\\
                 Institute for Nuclear Physics and J{\"u}lich Center for Hadron Physics,\\
                 Research Center J{\"u}lich, D-52425 J{\"u}lich, Germany\end{center}
}}
\maketitle
\begin{abstract}
Measurements of the $pp\to ppK^+K^-$ reaction, performed near the
kinematical threshold with the experiment COSY-11 at the Cooler Synchrotron COSY,
reveal a significant discrepancy between obtained excitation function and theoretical
expectations neglecting interactions of kaons.
In order to deepen our knowledge about the low energy dynamics
of the $ppKK$ system we investigated population of events for the
$pp\to ppK^+K^-$ reaction as a function of the invariant masses of two particle
subsystems. Based for the first time on the low-energy $K^+K^-$ invariant mass
distributions and the generalized Dalitz plot analysis, we estimated the scattering
length for the $K^+K^-$ interaction.
\end{abstract}
\section{Introduction}
The basic motivation for investigation of the $pp\to ppK^+K^-$ reaction near the kinematical 
threshold at COSY was an attempt to understand the nature of the scalar resonances
$f_{0}$(980) and $a_{0}$(980).
In addition to the standard interpretation as
$q\bar{q}$ states~\cite{Morgan}, these particles were also proposed to be $qq\bar{q}\bar{q}$
tetraquarks~\cite{Jaffe}, $K\bar{K}$ molecules~\cite{Lohse,Weinstein}, hybrid $q\bar{q}$/meson-meson
systems~\cite{Beveren} or even quark-less gluonic hadrons~\cite{Johnson}.
With regard to the formation of the molecule the strength of the $K\bar{K}$ interaction
becomes a crucial quantity, and it can be probed for example in the near threshold $pp\to ppK^+K^-$
reaction.
First measurements of this reaction were conducted at cooler synchrotron COSY by the
COSY-11 collaboration~\cite{wolke,quentmeier}. A precise determination of the collision energy, in the order of
fractions of MeV, permitted us to deal with  the rapid growth of cross sections~\cite{review} and thus to
take advantage of the threshold kinematics like
full phase space coverage achievable with dipole magnetic spectrometer being rather limited in geometrical acceptance.
These experiments revealed, however, that the total cross section at threshold is by more
than seven orders of magnitude smaller than the total proton-proton production
cross section making the study difficult due to low statistics.
A possible influence from the $f_{0}$ or $a_{0}$ on the $K^{+}K^{-}$ pair production
appeared to be too weak to be distinguished from the direct production of these mesons
on the basis of the COSY-11 data~\cite{quentmeier}.
However, the combined systematic collection of  data obtained by the collaborations
COSY-11~\cite{wolke,quentmeier,winter}, ANKE~\cite{anke} and DISTO~\cite{disto} reveal a significant signal
in the shape of the excitation function which  may be a manifestation of the interaction 
among particles in the final state.
\section{Total cross sections for the $pp\to ppK^+K^-$ reaction\\ near threshold}
\begin{figure}
\centering
\includegraphics[width=0.6\textwidth,angle=0]{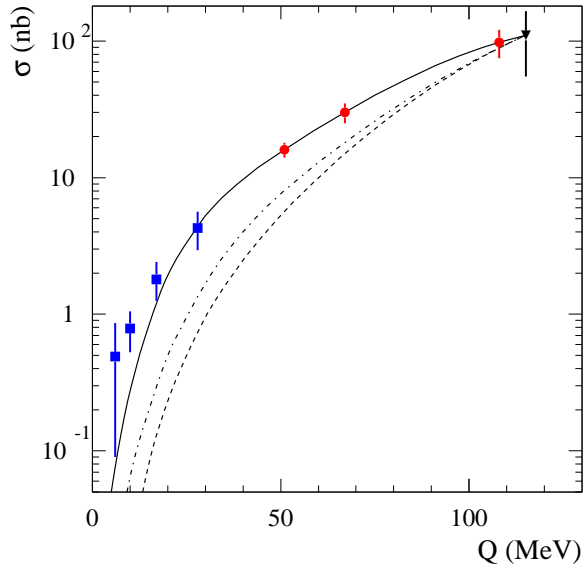}
\caption{Excitation function for the $pp\rightarrow ppK^{+}K^{-}$ reaction.
Triangle and circles represent the DISTO and ANKE measurements, respectively.
The four points close to the threshold are results from the COSY-11 measurements.
The curves are described in the text.}
\label{excitation-f}
\end{figure}

Results of all the measurements are presented in Fig.~\ref{excitation-f} 
together with curves representing three different theoretical expectations
normalized to the DISTO data point at Q~=~114~MeV~\cite{anke}.
The dashed curve represents the energy dependence from four-body phase space when we assume that
there is no interaction between particles in the final state. These calculations differ
by two orders of magnitude form data at Q~=~10~MeV and by a factor of
about five at Q~=~28~MeV.
Inclusion of the $pp$--FSI (dashed-dotted line in Fig.~\ref{excitation-f}), using parametrization
known from the three body final state~\cite{pp-FSI} with the four
body phase space, is closer to the experimental data, but does not fully
account for the difference~\cite{winter}. The enhancement may be due to the influence of
$pK$ and $K^{+}K^{-}$ interaction which was neglected in the calculations. Indeed,
the inclusion of the $pK^{-}$--FSI (solid line) reproduces the experimental data for
excess energies down to Q~=~28~MeV. These calculations of the cross section were accomplished
under the assumption that the overall enhancement factor, originating from final state
interaction in the $ppK^{+}K^{-}$ system, can be factorised into enhancements in the
$pp$ and two $pK^{-}$ subsystems~\cite{anke}:
 \begin{equation}
F_{FSI}~=~F_{pp}(q)\cdot F_{p_{1}K^{-}}(k_{1})\cdot F_{p_{2}K^{-}}(k_{2})~,
\label{pp-pkfsi}
\end{equation}
where $k_1$, $k_2$ and $q$ stand for the relative momenta of the particles in the first $pK^-$
subsystem, second $pK^-$ subsystem and $pp$ subsystem, respectively. Factors describing
the enhancement originating from the $pK^{-}$--FSI are parametrized using the scattering
length approximation, with the $pK^-$ scattering length amounting to $a_{pK^-} = (0 + 1.5i)$
fm~\cite{anke}. However the inclusion of the $pp$ and $pK^{-}$ final state interaction
fail to describe the data very close to threshold (see Fig.~\ref{excitation-f}).
This indicates that in this energy region the influence of the $K^{+}K^{-}$ interaction
is significant and cannot be neglected\footnote
{
In this calculations also the $pK^{+}$ interaction was neglected.
It is repulsive and weak and hence it can be 
interpreted as an additional attraction in the $pK^{-}$ system~\cite{anke}.
}.
Therefore we decided to perform more detailed analysis of the COSY-11 data
at excess energies of Q~=~10~MeV and 28~MeV including studies of both the differential cross
section distributions~\cite{ActaKK} and the strength of the final state interaction between the $K^{+}$ and
$K^{-}$~\cite{PhysRevC80_sil}.
\section{Analysis of the $K^+K^-$ final state interaction}
The final state interaction may manifest itself even stronger in the
distributions of the differential cross sections than in the shape of the excitation
function~\cite{review}. Thus, we have performed an analysis of the generalized Dalitz
plots~\cite{PhysRevC80_sil,nyborg} for the low energy data at Q~=~10~MeV (27~events)
and Q~=~28~MeV (30~events), in spite of the quite low statistics available.
\begin{figure}
\centering
\includegraphics[width=0.35\textwidth,angle=0]{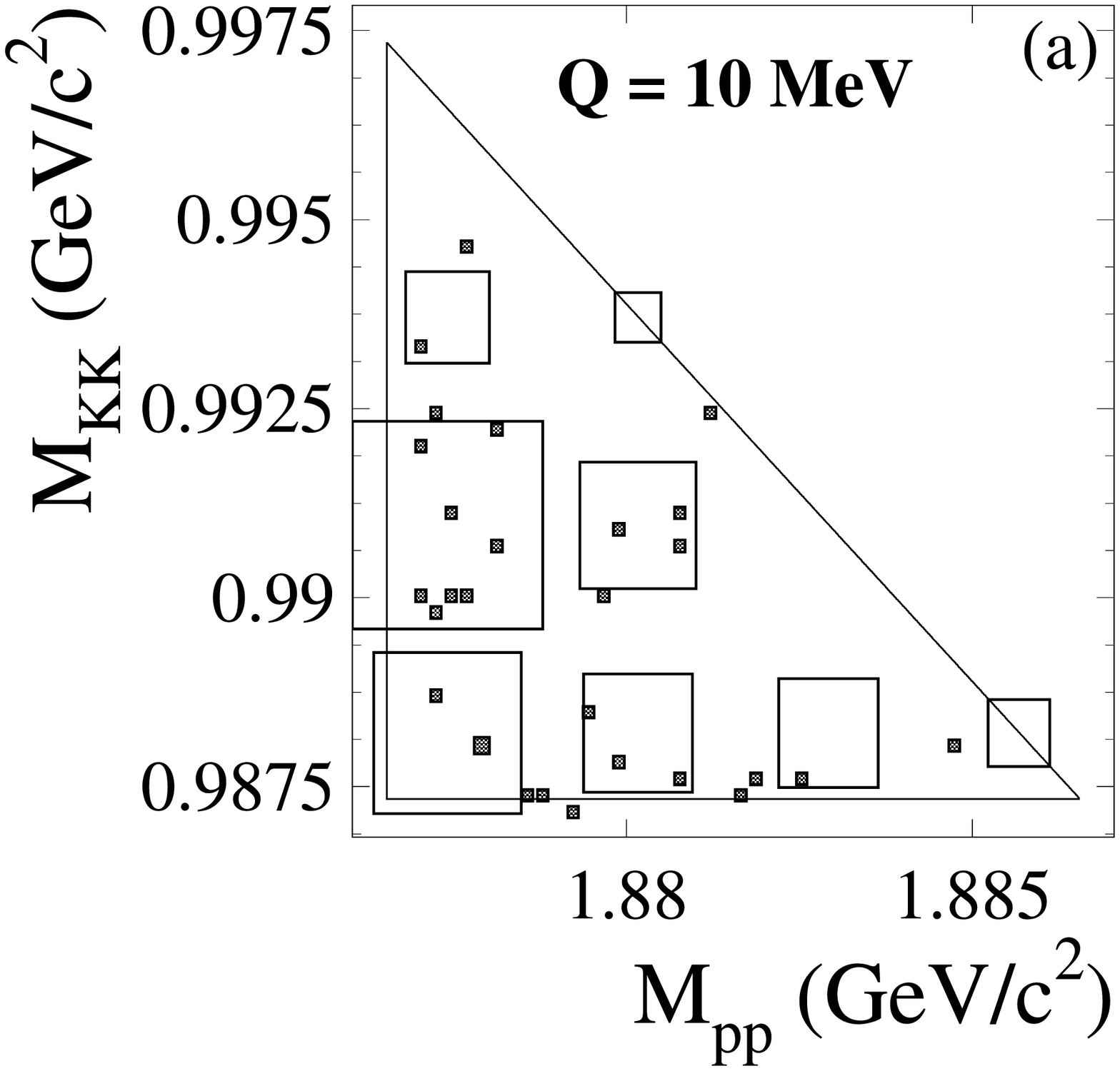}
\includegraphics[width=0.35\textwidth,angle=0]{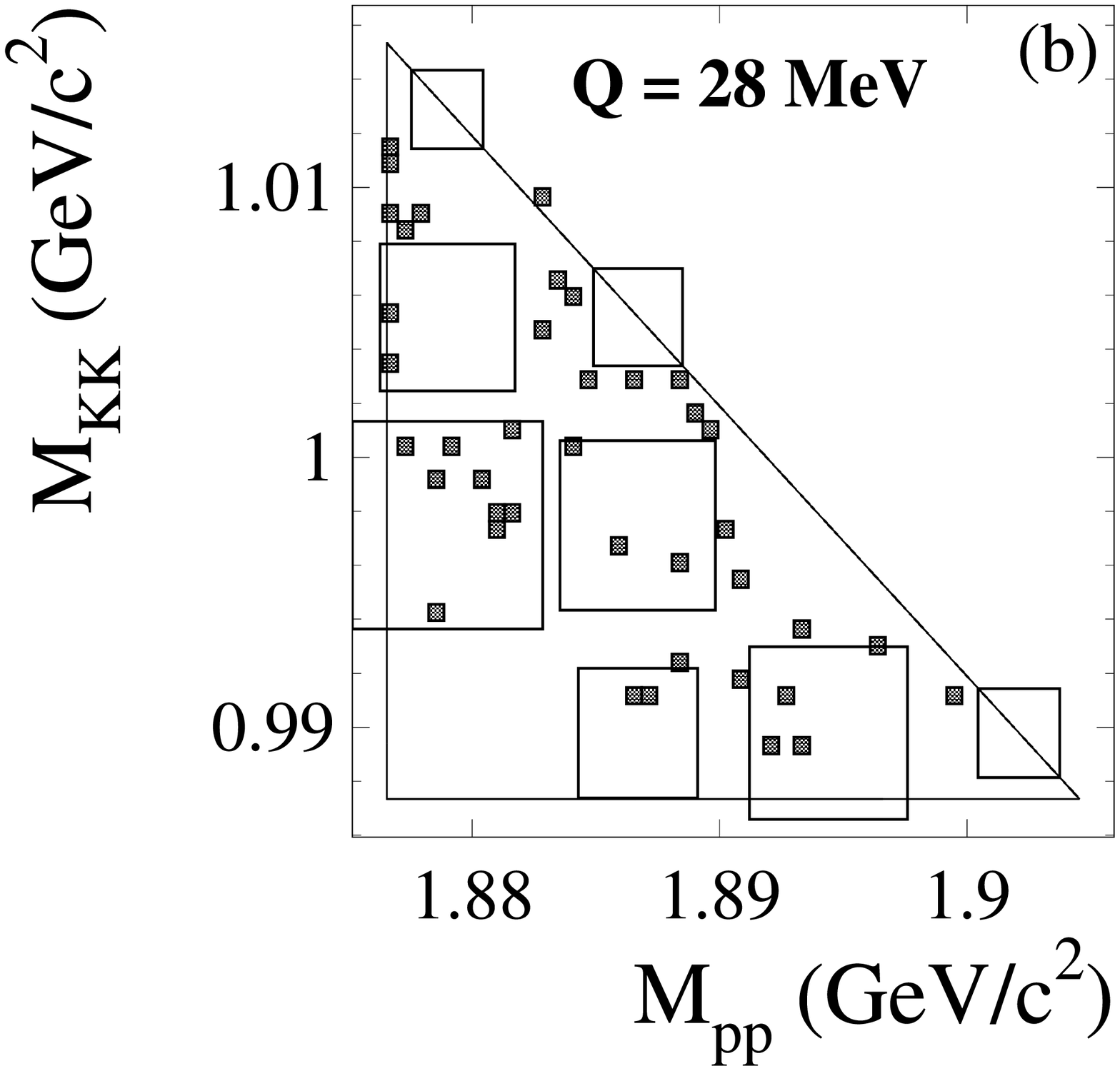}
\caption{
Goldhaber plots for the $pp\rightarrow ppK^{+}K^{-}$ reaction.
The solid lines of the triangles show the kinematically allowed boundaries.
Raw data are shown in Figs. (a) and (b)  as black points.
The superimposed squares represent the same distributions but binned into intervals of
$\Delta$M~=~2.5~MeV/c$^{2}$ ($\Delta$M~=~7~MeV/c$^{2}$) widths for an excess energy 
of Q~=~10~MeV (28~MeV), respectively.
The size of the square is proportional to the number of entries in a given interval.
}
\label{goldhabery}
\end{figure}
Complementary to previous derivations~\cite{kaminski,Baru,Teige,Bugg} here we estimate the $K^{+}K^{-}$
scattering length directly from the low energy differential mass distributions of $K^{+}K^{-}$ and pp pairs
from the $ppK^{+}K^{-}$ system produced at threshold.
The raw data (represented by black points in Figs.~\ref{goldhabery}(a) and ~\ref{goldhabery}(b)) were first binned
and then for each bin corrected for the acceptance and detection efficiency of the COSY-11 facility~\cite{mich_mgr}.
The resulting Goldhaber plots are presented together with the raw distributions 
in Figs.~\ref{goldhabery}(a) and \ref{goldhabery}(b).
In order to estimate the strength of the $K^+K^-$ interaction,
the derived cross sections were compared to results of simulations
generated with various parameters of the $K^{+}K^{-}$ 
interaction taking into account strong final state interaction
in the $pp$ and $pK^{-}$ subsystems.
To describe the experimental data in terms of final state interactions between
i) the two protons, ii) the $K^-$ and  protons and iii) the $K^{+}$ and $K^{-}$, the $K^{+}K^{-}$ enhancement
factor was introduced such that Eq.~\ref{pp-pkfsi} changes to:
\begin{equation}
F_{FSI}~=~F_{pp}(q)\cdot F_{p_{1}K^{-}}(k_{1})\cdot F_{p_{2}K^{-}}(k_{2})\cdot F_{K^{+}K^{-}}(k_{3})~.
\label{pp-pk-kk_fsi}
\end{equation}
As for the case of the $pK^{-}$--FSI, the $F_{K^+K^-}$ was calculated in the scattering length approximation:
\begin{equation}
F_{K^{+}K^{-}}~=~\frac{1}{1~-~i~k_{3}~a_{K^+K^-}}~,
\label{F_KK}
\end{equation}
where $a_{K^{+}K^{-}}$ is the effective $K^{+}K^{-}$ scattering length and $k_{3}$
stands for the relative momentum of the kaons in their rest frame.
Using this parametrization we compared the experimental event distributions
to the results of Monte Carlo simulations treating the $K^+K^-$ scattering length as
an unknown parameter, which has to be determined. In order to estimate the real and imaginary part of
$a_{K^{+}K^{-}}$ we constructed the Poisson likelihood $\chi^{2}$ statistic derived from the
maximum likelihood method~\cite{bakernim,feldmanpr}. Data collected at both excess energies have been
analysed simultaneously~\cite{PhysRevC80_sil}. The best fit to the experimental data corresponds
to $\left|Re(a_{K^{+}K^{-}})\right| = 0.5^{~+4}_{~-0.5}$~fm and $Im(a_{K^{+}K^{-}}) = 3~\pm~3$~fm.
The final state interaction enhancement factor $F_{K^{+}K^{-}}$ in the scattering length approximation
is symmetrical with respect to the sign of $Re(a_{K^{+}K^{-}})$,
therefore only its absolute value can be determined.
\section{Summary}
The analysis of the $pp\rightarrow ppK^{+}K^{-}$ reaction measured by COSY-11 collaboration
at excess energy Q = 10 MeV and Q = 28 MeV  has been extended to the
determination of the differential cross sections in view of the $K^{+}K^{-}$ final state interaction.
The extracted $K^+K^-$ scattering length amounts to:
\begin{center}
$\left|Re(a_{K^+K^-})\right|$~=~0.5$^{~+4}_{~-0.5}$~fm\\
$Im(a_{K^+K^-})$~=~3~$\pm$~3~fm.
\end{center}
Due to the low statistics the uncertainties are rather large.
In this analysis we cannot distinguish between the isospin I~=~0 and I~=~1 states of the $K^+K^-$ system.
However, as pointed out in~\cite{dzyuba}, the production with I~=~0 is dominant in the $pp\to ppK^+K^-$ reaction
independent of the exact values of the scattering lengths.


\begin{thebibliography}{99}
\bibitem{Morgan}
D.~Morgan,~M.~R.~Pennington,~Phys. Rev.~D~\textbf{48}, 1185 (1993).
\bibitem{Jaffe}
R.~L.~Jaffe,~Phys. Rev.~D \textbf{15}, 267 (1977).
\bibitem{Lohse}
D.~Lohse~\textit{et al.},~Nucl. Phys. \textbf{A516}, 513 (1990).
\bibitem{Weinstein}
J.~D.~Weinstein,~N.~Isgur,~Phys. Rev.~D~\textbf{41}, 2236 (1990).
\bibitem{Beveren}
E.~Van Beveren,~\textit{et al.}, Z. Phys.~C~\textbf{30}, 615 (1986).
\bibitem{Johnson}
R.~L.~Jaffe,~K.~Johnson,~Phys. Lett. \textbf{B60}, 201 (1976).
\bibitem{wolke}
M.~Wolke,~PhD thesis, IKP J{\"u}l-3532 (1997).
\bibitem{quentmeier}
C.~Quentmeier~\textit{et al.},~Phys. Lett. \textbf{B515}, 276 (2001).
\bibitem{review}
P.~Moskal,~\textit{et al.}, Prog. Part. Nucl. Phys. \textbf{49}, 1 (2002).
\bibitem{winter}
P.~Winter~\textit{et al.},~Phys. Lett. \textbf{B635}, 23 (2006).
\bibitem{anke}
Y.~Maeda~\textit{et al.},~Phys. Rev.~C \textbf{77}, 01524 (2008).
\bibitem{disto}
F.~Balestra~\textit{et al.},~Phys. Lett. \textbf{B468}, 7 (1999).
\bibitem{pp-FSI}
P.~Moskal~\textit{et al.},~Phys. Lett. \textbf{B482}, 356 (2000).
\bibitem{ActaKK}
M.~Silarski,~\textit{et al.}, Acta Phys.\ Polon.\ Supp.\ {\bf 2} 97 (2009).
\bibitem{PhysRevC80_sil}
M.~Silarski~\textit{et al.},~Phys. Rev.~C \textbf{80}, 045202 (2009).
\bibitem{nyborg}
P.~Nyborg~\textit{et al.},~Phys. Rev. \textbf{140}, 914 (1965).
%
\bibitem{kaminski}
R.~Kami\'nski, L. Le\'sniak,~Phys. Rev.~C \textbf{54}, 2264 (1995).
\bibitem{Baru}
V.~Baru~\textit{et al.}, Phys. Lett.~\textbf{B586}, 53 (2004).
\bibitem{Teige}
S.~Teige~\textit{et al.}, Phys. Rev.~D \textbf{59}, 012001 (2001).
\bibitem{Bugg} 
D. V.~Bugg,~\textit{et al.}, Phys. Rev.~D \textbf{50}, 4412 (1994).
\bibitem{mich_mgr}
M.~Silarski, FZ-J\"ulich report, J\"ul-4278, (2008).
\bibitem{bakernim}
S. Baker, R.D. Cousins,~Nucl. Instrum. Methods Phys. Res.~A \textbf{221}, 437 (1984).
\bibitem{feldmanpr}
  G.~J. Feldman, R.~D. Cousins, Phys. Rev.~D \textbf{57}, 3873 (1998).
\bibitem{dzyuba}
A. Dzyuba~\textit{et al.}, Phys. Lett.~\textbf{B668}, 315 (2008).
\end{thebibliography}
\end{document}